\documentclass{PoS}

\usepackage{wrapfig}
\usepackage{dcolumn}
\usepackage{latexsym}

\newcommand{\be}{\begin{equation}}
\newcommand{\ee}{\end{equation}}
\newcommand{\bea}{\begin{eqnarray}}
\newcommand{\eea}{\end{eqnarray}}

\newcommand{\refc}[1]{(\ref{#1})}
\newcommand{\Ord}{O}

\newcommand{\bt}{\bar{b}_2}
\newcommand{\btt}{$\bt$}

\newcommand{\SU}{{\rm SU}}
\newcommand{\U}{{\rm U}}

\title{Spectrum of the open QCD flux tube and its effective string
description\thanks{I would like to thank the conveners of Section A "Vacuum structure and
confinement" for the invitation to present my results. I am also
grateful to Dmitri Antonov, Cristina Diamantini, Stefan Floerchinger and
Edward Shuryak for stimulating discussions and Francesca Cuteri for a careful reading
of the manuscript. This research has been funded by
the DFG via the Emmy Noether Programme EN 1064/2-1 and SFB/TRR 55. I have
also received support from the Frankfurter F\"orderverein f\"ur Physikalische
Grundlagenforschung.}}

\ShortTitle{Spectrum of the open QCD flux tube and its effective string description}

\author{\speaker{Bastian B. Brandt}\\
        Institute for Theoretical Physics, Goethe University, Max-von-Laue-Strasse 1, 60438 Frankfurt am Main, Germany \\
        E-mail: \email{brandt@th.physik.uni-frankfurt.de}}

\abstract{I perform a high precision measurement of the static quark-antiquark potential in
three-dimensional $\SU(N)$ gauge theory with $N=2$ to 6. The results are compared to the effective
string theory for the QCD flux tube and I obtain continuum limit results for the string tension and
the non-universal leading order boundary coefficient, including an extensive analysis of all types of
systematic uncertainties. The magnitude of the boundary coefficient decreases with increasing $N$, but
remains non-vanishing in the large-$N$ limit. I also test for the presence of possible contributions
from rigidity or massive modes and compare the results for the string theory parameters to data for
the excited states.}

\FullConference{XIII Quark Confinement and the Hadron Spectrum - Confinement2018\\
		31 July - 6 August 2018\\
		Maynooth University, Ireland}

\begin{document}

\section{Introduction}

The observation that the lowest lying mesons can be grouped in so-called Regge-trajectories has lead
to the formulation of the first string theories~\cite{Goto:1971ce,Goddard:1973qh} to describe the strong
interactions. This string picture even persists until today, after the introduction of Quantum
Chromodynamics (QCD) as the fundamental theory for the strong interactions, in terms of an effective
string theory (EST) for mesons and baryons. The basic idea is that, for large quark distances,
the chromo-electromagnetic field connecting the quarks is squeezed into a narrow, tube-alike region,
a flux tube, via a dual version of the Meissner effect. For a quark-antiquark ($q\bar{q}$) pair, i.e.,
a mesonic state, this is shown
\begin{wrapfigure}{r}{7cm}
 \centering
 \includegraphics[width=5.55cm]{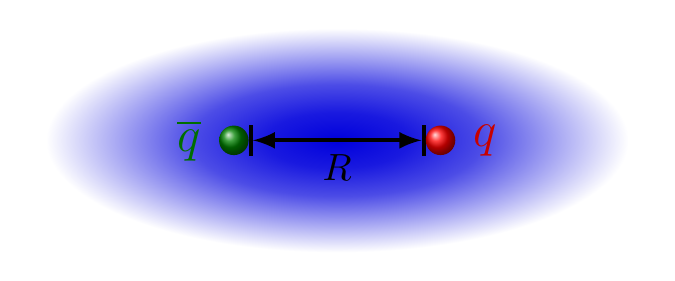}\\[2mm]
 $\Big\Downarrow$ enlarging $R$\\[2mm]
 \includegraphics[width=7cm]{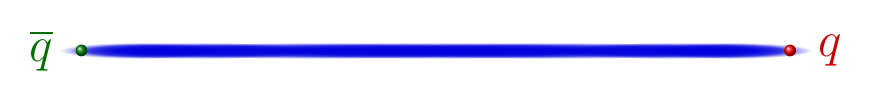}
 \caption{\label{fig:flux-tube}
 Schematic picture of flux tube formation for a static $q\bar{q}$ pair.
 }
\end{wrapfigure}
schematically in Fig.~\ref{fig:flux-tube}. As soon as the width of the
flux tube becomes negligible compared to the $q\bar{q}$ separation $R$, the flux tube effectively looks like
a thin energy string (also denoted as the confining string in this framework) and its dynamics
is governed by stringy excitations described by the EST.
The resulting dominant term in the energy levels at large $R$ is a linear term of the form $\sigma R$, where
$\sigma$ is the energy per unit length of the flux tube, known as the string tension.
Consequently, flux tube formation provides a heuristic mechanism to explain quark confinement in QCD
(see Ref.~\cite{Greensite:2003bk} for a review).
In QCD with finite quark masses, the flux tube persists only up to $q\bar{q}$ distances where the potential
energy in the system allows to create another $q\bar{q}$ pair from the vacuum, leading to a state with
two instead of one mesons. This effect is known as string breaking (see Ref.~\cite{Bali:2005fu} for a
study in full QCD, for instance) and it is the reason why quarks have not been observed as free particles in
nature.

In the static limit, the energy of a $q\bar{q}$ pair at distance $R$ is related to the static 
quark-antiquark potential $V(R)$. Similar potentials can also be defined for mesonic states with
excited gluon configurations carrying non-trivial quantum numbers. These potentials are, besides their
relevance concerning the anatomy of confinement, important
for the theoretical description of heavy quarkonia and hybrid
mesons.\footnote{See Refs.~\cite{Brambilla:2010cs,Meyer:2015eta} for reviews and
Ref.~\cite{Capitani:2018rox} for a recent lattice study of hybrid mesons.} To study these systems
the availability of an analytic expression for the potentials is important and the EST can provide
valuable input.

\begin{figure}[t]
 \centering
 \includegraphics[width=9cm]{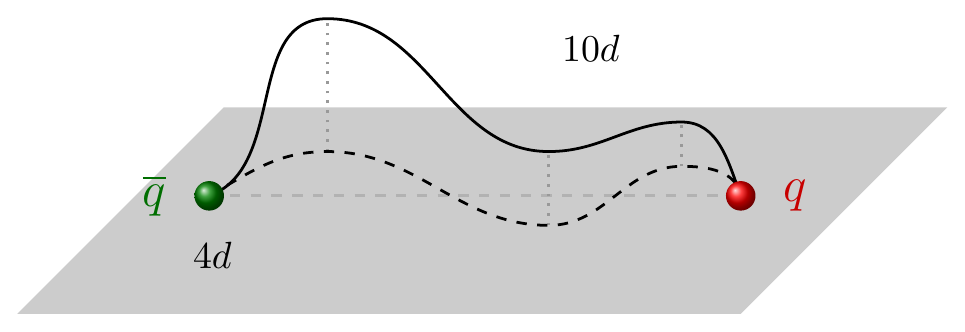}
 \caption{\label{fig:adscft}
 Schematic picture of the projection of the 10-dimensional string in a generalization of the AdS/CFT
 correspondence to large-$N$ gauge theories onto the 4-dimensional boundary of the AdS space. The
 boundary is associated with the spacetime of pure gauge theory.
 }
\end{figure}

Apart from these phenomenological applications, the EST provides a natural framework to make contact
with a possible 10-dimensional (10d) string theory dual to large-$N$ gauge theories in terms of a
generalization of the AdS/CFT correspondence~\cite{Maldacena:1997re}. In principle, the EST
action can be computed from the 10d string theory by integrating out the additional massive
modes (see Ref.~\cite{Aharony:2009gg} and references therein). In terms of this correspondence the confining
string can be visualized, as shown in Fig.~\ref{fig:adscft}, as the 4d projection of the 10d string
onto the boundary of the anti-de Sitter space, associated with the spacetime of the gauge theory.
When computed from the fundamental string theory, possible non-universal
parameters in the EST action are related to properties of the AdS side of the duality.
A particular example is the boundary coefficient \btt{}, introduced in the next section, which,
for certain holographic string backgrounds, can be related to the masses of the additional bosonic
and fermionic degrees of freedom of the fundamental string theory~\cite{Aharony:2010cx}.
Knowledge about the non-universal parameters in the $N\to\infty$ limit can thus be converted into
constraints for the 10d string theory in AdS spacetime and help to find
suitable backgrounds for the gauge/gravity correspondence of non-conformal gauge theories. See
Refs.~\cite{Kol:2010fq,Vyas:2012dg,Giataganas:2015yaa} for recent computations
of properties of flux tubes in this framework.

The energy levels of the flux tube and the associated potentials can be computed in pure gauge
theory on the lattice. A recent collection of results can be found in Ref.~\cite{Brandt:2016xsp}.
All of the results so far have shown remarkable agreement with the predictions from the EST
down to comparably small values of $\sqrt{\sigma}R\approx 1.2 - 1.5$. Note, that the EST, as an
effective theory for long flux tubes, is expected to break down at around
$\sqrt{\sigma}R\lesssim 1$.
In 4d, energy states of closed flux tubes have been found recently~\cite{Athenodorou:2010cs}
which are consistently described by massive modes on the string
worldsheet~\cite{Dubovsky:2013gi,Dubovsky:2014fma}. Open string states with a similar behavior
have also been observed earlier~\cite{Juge:2002br}. The appearance of such states is certainly
expected, given the differences between the flux tube and a string. The finite width of the flux tube
may allow for inner vibrations and torsions, which could show up as massive modes in the spectrum.
A peculiar feature is the apparent absence of massive modes in the 3d data. A theoretical explanation
for this is the fact that the coupling term between the massive mode and the Goldstone bosons does not
exist in 3d. From the phenomenological point of view, neither torsions nor the development of knots are
possible, since there is only one transverse direction. Thus the absence of the massive modes in 3d
is plausible if the low lying massive modes are related to either of these phenomena.
In 3d we are left with vibration modes and contributions due to the
string rigidity. The latter has been first proposed by Polyakov~\cite{Polyakov:1986cs} and recently
found to be essential to describe the potential in 3d $\U(1)$ gauge theory~\cite{Caselle:2014eka}.
Rigidity contributions appear at high orders in the $1/R$ expansion, but the presence of the rigidity
term gives a non-perturbative contribution to the
potential~\cite{Klassen:1990dx,Nesterenko:1997ku,Caselle:2014eka}. Formally its contribution
is similar to the one of a free massive mode on the worldsheet (coupling to the
Goldstone bosons only via the induced metric).

In this proceedings article I report on the progress concerning my
studies~\cite{Brandt:2009tc,Brandt:2010bw,Brandt:2013eua,Brandt:2017yzw}
of the energy levels of the flux tube and their comparison to the EST in
3d $\SU(N)$ gauge theories for $N\to\infty$. I will show continuum extrapolated
results for the string tension (comparing also to the Karabali-Kim-Nair
(KKN) prediction~\cite{Karabali:1998yq}) and the boundary coefficient \btt{} for $N=2$ to 6,
which are then extrapolated to the large-$N$ limit. \btt{} is first obtained from an analysis
excluding massive modes or contributions from rigidity. In the second step
we test whether such modes can be present, how they change the value of \btt{} and extract their mass.

\section{EST predictions and massive/rigid modes}

The EST describes the dynamics of a stable non-interacting flux tube in terms of the quantized
transverse oscillation modes, the Goldstone bosons associated with the breaking of the
translational symmetry by the tube. The basic properties of the theory are known for some
time~\cite{Nambu:1978bd,Luscher:1980fr,Polyakov:1980ca} but a number of features have only been elucidated
recently (for reviews see~\cite{Aharony:2013ipa,Brandt:2016xsp}). The spectrum has been computed up to
$O(R^{-5})$~\cite{Aharony:2010db,Aharony:2011ga},
\be
\label{eq:est-spec}
\begin{array}{rl}
\displaystyle E^{\rm EST}_{n,l}(R) & \displaystyle =
\sigma \: R \: \sqrt{ 1 + \frac{2\pi}{\sigma\:R^{2}} \:
\left( n - \frac{1}{24} \: ( d - 2 ) \right) } \vspace*{2mm} \\
 & \displaystyle - \bt \frac{\pi^3}{\sqrt{\sigma^3} R^4}
\Big( B_n^l + \frac{d-2}{60} \Big) - \frac{\pi^3 (d-26)}{48 \sigma^2 R^5}
C_n^l + \Ord(R^{-\xi}) \,.
\end{array}
\ee
The leading order term, the first term on the right-hand-side, is the spectrum obtained from the
light cone quantization~\cite{Arvis:1983fp} of the Nambu-Goto string
(LC spectrum). $B_n^l$ and $C_n^l$ are dimensionless coefficients, depending on the representation
of the state with respect to rotations around the string axis (see~\cite{Brandt:2017yzw}).
The term proportional to \btt{} is the leading order
boundary correction (BC). For the groundstate in 3d $B_0^0$ and $C_0^0$
vanish, so that the BC is the only correction term up to $\Ord(R^{-6})$. \btt{} has
previously been computed in 3d $\SU(2)$~\cite{Brandt:2010bw} and $Z(2)$ gauge
theories~\cite{Billo:2012da}.

In the energy levels of Eq.~\refc{eq:est-spec}, possible contributions from rigidity or massive
modes have not been included. In 3d contributions from rigidity and
massive bosons on the worldsheet are formally equivalent. In $\zeta$-function regularization and
including higher order terms perturbatively, the rigidity/massive mode corrections are given
by~\cite{Caselle:2014eka}
\be
\label{eq:pot-rigid}
V^{\rm rig}(R) = - \frac{m}{2\pi} \sum_{k=1}^{\infty} \frac{K_1(2kmR)}{k}
- \frac{(d-2)(d-10)\pi^2}{3840 m \sigma R^4} \,.
\ee
Here $m$ is the mass parameter, which is related to the rigidity parameter in case
of a correction originating from the string rigidity, and $K_1$ is a modified Bessel functions
of the second kind. The second term on the right-hand-side
contaminates the BC term and thus changes the value of \btt{}. Note, that the $\zeta$-function
regularization scheme breaks Lorentz symmetry, so that counterterms may need to
be taken into account for a proper extraction of the energy levels~\cite{Dubovsky:2012sh}.
From now on we will always refer to the correction terms from Eq.~\refc{eq:pot-rigid} as
``massive mode'' contributions.

\section{String tension and KKN prediction}
\label{sec:sigma}

We perform simulations in 3d $\SU(N)$ ($N=2$ to 6) gauge theory
using the standard mixture of heatbath and overrelaxation steps.
The potential is extracted from Polyakov loop correlation functions, which are computed with
one level of the L\"uscher-Weisz multilevel algorithm~\cite{Luscher:2001up} for error reduction.
We used 20000 sublattice updates and temporal sublattice sizes ranging between 2 and 12.
For more details and a study of systematic effects see Ref.~\cite{Brandt:2017yzw}. For scale
setting we use the Sommer scale $r_0=0.5$~fm~\cite{Sommer:1993ce}, which may be used to translate to
``physical'' units. For each value of $N$ we have simulated at least 3 lattice spacings,
keeping the spatial and temporal extents larger than $10 \, r_0$ to render finite size effects
negligible. The potential has been extracted up to $R/r_0\gtrsim3$ for all lattice spacings.

\begin{figure}[t]
 \centering
 \includegraphics[width=12cm]{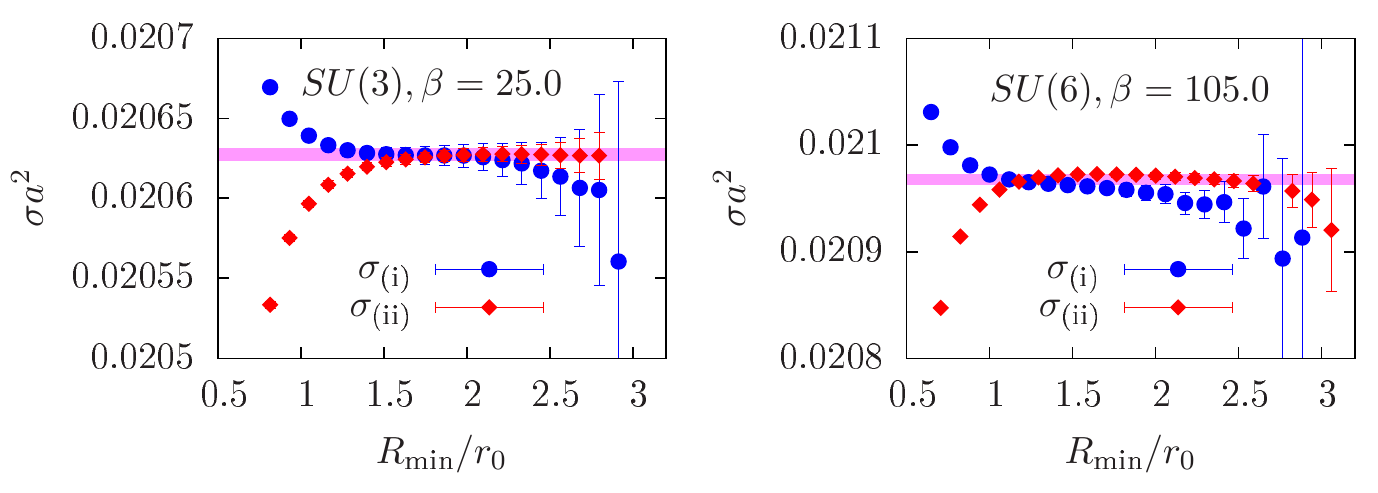}
 \caption{\label{fig:sig-extract}
 Results for the string tension extracted from fits (i), $\sigma_{\rm (i)}$,
 and (ii), $\sigma_{\rm (ii)}$, as explained in the text, versus
 $R_{\rm min}$ in units of $r_0$ for $N=3$ (left) and 6 (right). The magenta
 bands are the values for $\sigma_{(ii)}$ used in the further analysis.
 }
\end{figure}

\begin{figure}[t]
 \centering
 \includegraphics[width=12cm]{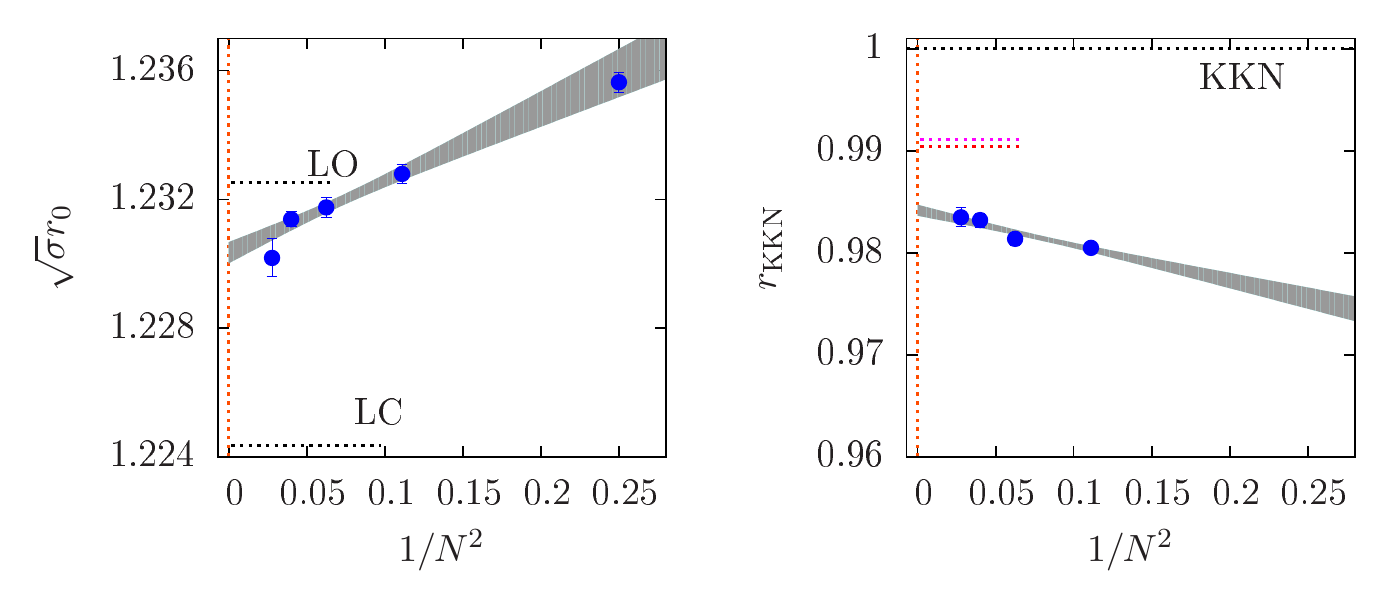}
 \caption{\label{fig:sig-largeN}
 Continuum results for $\sqrt{\sigma}$ (left) and the ratio $r_{\rm KKN}$
 (right) versus $1/N^2$. The curves are the large-$N$ extrapolations. The
 dashed line in the left plot labeled by `LC' indicates the values of
 $\sqrt{\sigma}r_0$ obtained from the full LC spectrum and the one labeled
 by `LO' is the  result from the expansion to $\Ord(1/R)$. The dashed lines
 in the right plot are the results for $r_{\rm KKN}$
 from~\cite{Lucini:2002wg} (red) and~\cite{Bringoltz:2006gp} (magenta).
 }
\end{figure}

We start by extracting the string tension $\sigma$ in a way that we can control the effect of
higher order corrections. To this end we use of two different fits: (i) we fit the force
($F(R)=\partial V/[\partial R]$) to the form $ \: R^2 F(R) = \sigma R^2 + \gamma $;
(ii) we fit $V(R)$ to the LC potential, adding a normalization constant $V_0$.
The fits include different terms of the $1/R$ expansion starting at
$O(1/R^3)$, so that the results will differ once corrections at this order become important.
To isolate the asymptotic behavior, one can thus investigate the dependence of the fit parameters
on the minimal value of $R$ included in the fit, $R_{\rm min}$. In the region where the results
from the fits agree within errors and show a plateau the estimate for $\sigma$ from either
of the methods will be reliable within the given uncertainties. Two typical examples for the
$R_{\rm min}$ dependence of $\sigma$ obtained from the two fits, denoted as $\sigma_{\rm (i)}$ and
$\sigma_{\rm (ii)}$, respectively, are shown in Fig.~\ref{fig:sig-extract}. In the
following we will always use the result $\sigma_{\rm (ii)}$ obtained with the value of
$R_{\rm min}$ where the results of the two methods become fully consistent (indicated by the magenta
bands in Fig.~\ref{fig:sig-extract}). We extrapolate $\sqrt{\sigma}r_0$ to the
continuum including terms of $\Ord(a^2)$ and $\Ord(a^4)$. The systematic uncertainty of this
extrapolation is estimated by comparison to a fit for the data with $r_0/a\gtrsim6$, including only
a term of $\Ord(a^2)$. The continuum results for $\sigma$ are shown in
Fig.~\ref{fig:sig-largeN} (left) versus $1/N^2$.

To compare to the KKN prediction~\cite{Karabali:1998yq},
\be
\label{eq:kkn}
\frac{\sqrt{\sigma}}{g_{\rm MF}^2} = \sqrt{\frac{N^2-1}{8\pi}} \,,
\quad \textnormal{with} \quad
g_{\rm MF}^2 = \frac{2N}{\beta \big< U_p \big>} \,,
\ee
the mean-field improved coupling~\cite{Lepage:1992xa}, and the plaquette
expectation value $\big< U_p \big>$, we define
\be
\label{eq:kkn-ratio}
r_{\rm KKN} = \frac{\big(\sqrt{\sigma}/g_{\rm MF}^2\big)_{\rm
 lat}}{\sqrt{(N^2-1)/(8\pi)}} \,,
\ee
where the denominator includes the lattice result. We compute the
ratio in the continuum using a continuum extrapolation of
$g_{\rm MF}^2$ in units of $r_0$ as for $\sigma$ with $a^2\to a$.
The results for the ratio are shown in Fig.~\ref{fig:sig-largeN} (right).

Finally, we extrapolate the results to $N\to\infty$ using a function linear
in $1/N^2$ excluding the data with $N=2$. The resulting extrapolations
are shown by the gray curves in both panels of Fig.~\ref{fig:sig-largeN}. To quantify
the systematic uncertainty of the extrapolation, we repeat the extrapolation
excluding the $N=3$ result. The results for $\sigma$ and $r_{\rm KKN}$ in
the large-$N$ limit are
\be
\label{eq:sig-r-res}
 r_0\sqrt{\sigma}^{N\to\infty} = 1.2304 (5) \quad \textnormal{and} \quad
 r_{\rm KKN}^{N\to\infty} = 0.9842 (15) \,.
\ee
The value for $\sqrt{\sigma}r_0$ is unambiguously determined
within the EST and we display the values for the LC spectrum and its leading
order (LO) expansion in $1/R$ by the dashed lines in Fig.~\ref{fig:sig-largeN}.
The large-$N$ extrapolation clearly lies between the two cases, indicating that
corrections to the LC spectrum are mandatory to describe the large-$N$ potential
down to $R=r_0$. In the right panel of Fig.~\ref{fig:sig-largeN} we also show
results for $r_{\rm KKN}$ from the
literature~\cite{Lucini:2002wg,Bringoltz:2006gp} (dashed lines).
Our results turn out to be somewhat smaller than the ones from previous
computations.

\section{EST analysis without massive modes}

\begin{figure}[t]
 \centering
 \includegraphics[width=12cm]{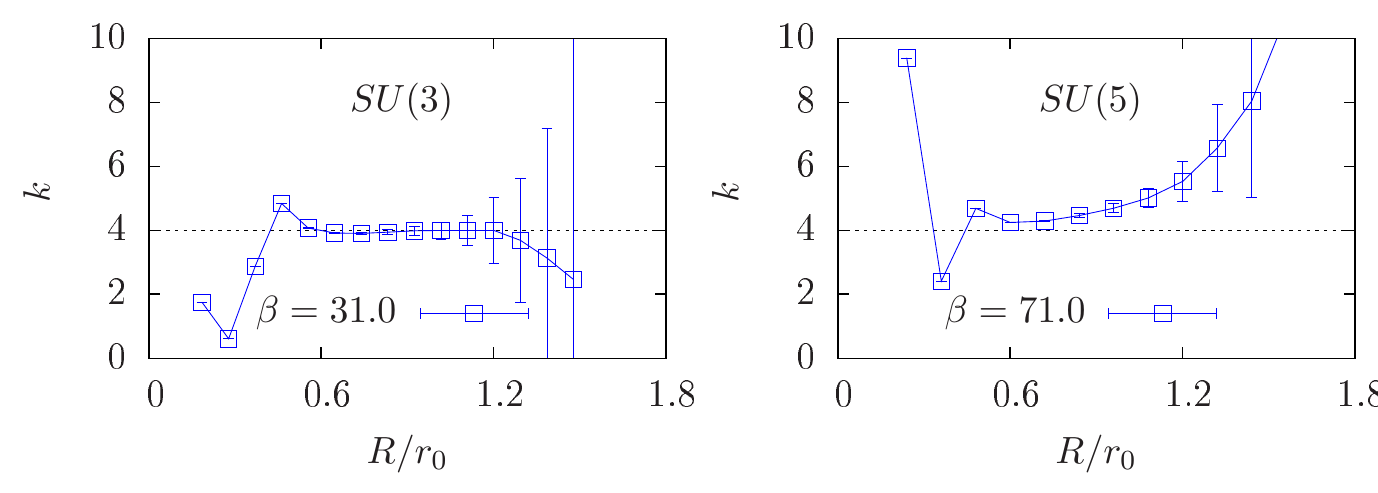}
 \caption{\label{fig:lead-plot}
 Examples for the dependence of the exponent $k$ on $R_{\rm min}$
 for $N=3$ (left) and 5 (right).
 }
\end{figure}

The EST predicts corrections to the LC spectrum starting at $O(R^{-4})$.
To test whether this is reflected by the data we fit them to the form
\be
\label{eq:lead-coeff-fit}
V(R) = \sigma \: R \: \sqrt{ 1 - \frac{\pi}{12\:\sigma\:R^{2}} }
+ \frac{\eta}{\big(\sqrt{\sigma} R\big)^k} + V_0 \,,
\ee
where $k$, $\eta$, $\sigma$ and $V_0$, are fit parameters. If the data
agree with the EST predictions, we expect to see a clear plateau at $k=4$
when $k$ is plotted versus $R_{\rm min}$. Two typical
examples for the $R_{\rm min}$ dependence of $k$ are shown
in Fig.~\ref{fig:lead-plot}. While the results for $\SU(3)$ indicate a clear
plateau at $k=4$, the $\SU(5)$ results show values of $k\gtrsim4$. This is
generally the case for larger $N$ and possibly indicates a decrease of
\btt{} (which we will indeed find below), so that higher order corrections
become more important with increasing $N$.

To extract \btt{}, at this point neglecting massive mode corrections, we
use the general fit formula
\be
\label{eq:boundary-fit}
V(R) = E^{\rm EST}_{0}(R) + \frac{\gamma^{(1)}_{0}}{\sqrt{\sigma^5} R^6} + 
\frac{\gamma^{(2)}_{0}}{\sigma^3 R^7} + V_0 \,.
\ee
Here $E^{\rm EST}_{0}$ is the energy level with $N,l=0$ from
Eq.~\refc{eq:est-spec} and $\gamma^{(1)}_{0}$, $\gamma^{(2)}_{0}$ and $V_0$
are fit parameters along with $\sigma$ and \btt. We perform five different fits:
\begin{enumerate}
 \vspace*{-2mm}
 \item[{\bf A}] take $\sigma$ and $V_0$ from Sec.~\ref{sec:sigma}, use \btt,
 $\gamma^{(1)}_{0}$ and $\gamma^{(2)}_{0}$ as free parameters;
 \vspace*{-2mm}
 \item[{\bf B}] use $\sigma$, $V_0$  and \btt{} as free parameters, set
 $\gamma^{(1)}_{0}=0$ and $\gamma^{(2)}_{0}=0$;
 \vspace*{-2mm}
 \item[{\bf C}] use $\sigma$, $V_0$,  \btt{} and $\gamma^{(1)}_{0}$ as free
 parameters, set $\gamma^{(2)}_{0}=0$;
 \vspace*{-2mm}
 \item[{\bf D}] use $\sigma$, $V_0$, \btt{} and $\gamma^{(2)}_{0}$ as free
 parameters, set $\gamma^{(1)}_{0}=0$;
 \vspace*{-2mm}
 \item[{\bf E}] use $\sigma$, $V_0$, $\gamma^{(1)}_{0}$ and $\gamma^{(2)}_{0}$
 as free parameters, set $\bt=0$.
 \vspace*{-2mm}
\end{enumerate}
For all of the fits we use the second smallest value for $R_{\rm min}$ for which
$\chi^2/$dof$<1.5$. Better or worse agreement with the data is then indicated by
the value of $R_{\rm min}$ in combination with the number of higher order terms
included in the fit. For fit {\bf C}, for instance, we expect a smaller value of
$R_{\rm min}$ compared to fit {\bf B}, which does not contain higher order
corrections. For all parameter sets fit {\bf E} demands larger values for
$R_{\rm min}$ compared to fits {\bf C} and {\bf D}, indicating less agreement
with the data. This is possibly due to the fact that both correction terms are
needed to mimic the $1/R^4$ term at intermediate distances, which shows that
the $1/R^4$ term is necessary to successfully describe the data. For fit {\bf A},
$R_{\rm min}$ has to be larger than for fits {\bf C} and {\bf D}, showing
that it is too restrictive to fix the values of $\sigma$ and $V_0$ in the fits,
even though the change of $\sigma$ is not significant. In the following we thus
use fits {\bf B} to {\bf D} and their weighted average as the final result.

\begin{figure}[t]
 \centering
 \includegraphics[width=12cm]{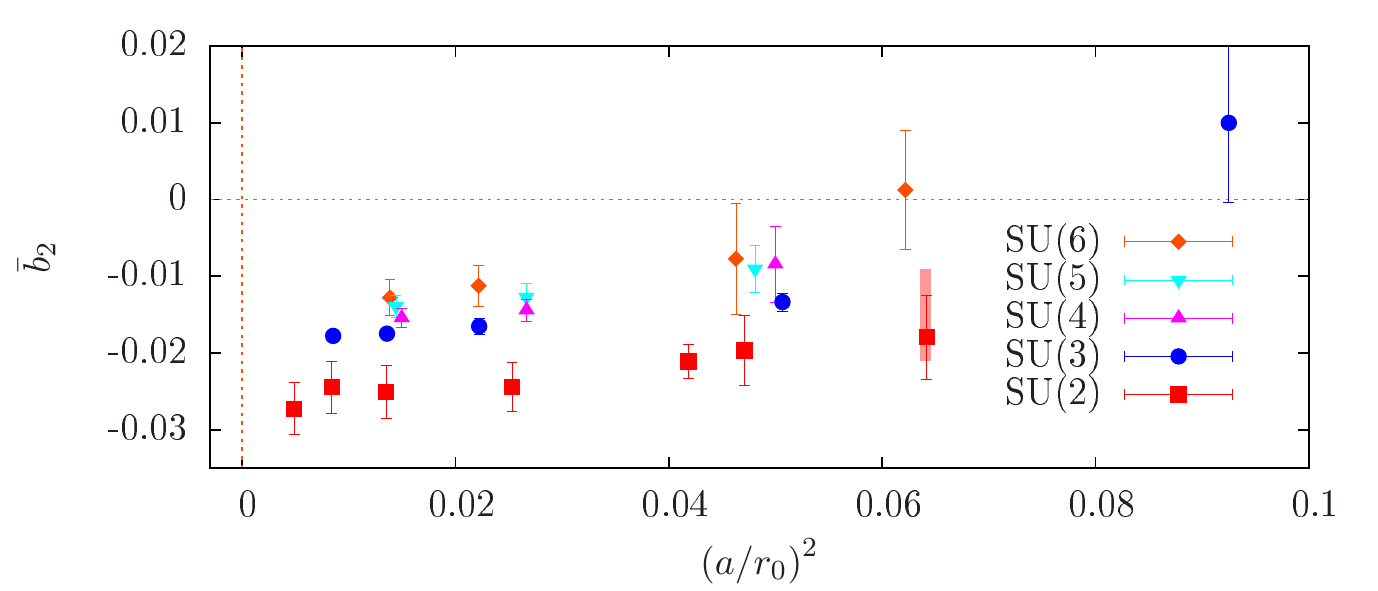}
 \caption{\label{fig:b2vsasq}
 Results for the boundary coefficient \btt{} for the individual lattices versus
 the squared lattice spacing in units of $r_0$. The red band indicates the result
 for \btt{} obtained from the excited states in~\cite{Brandt:2010bw}.
 }
\end{figure}

\begin{figure}[t]
 \centering
 \includegraphics[width=12cm]{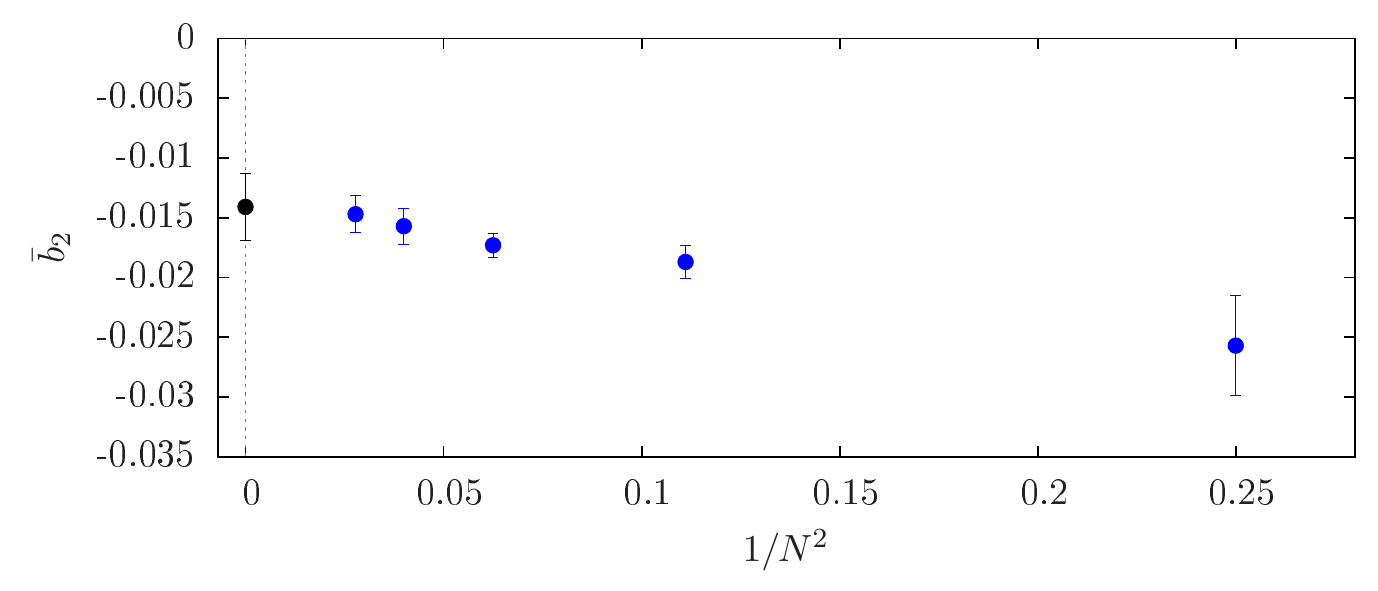}
 \caption{\label{fig:b2vsNsq}
 Continuum results for the boundary coefficient \btt{} versus $1/N^2$. The black
 point is the result in the large-$N$ limit.
 }
\end{figure}

The results for \btt{} are shown in Fig.~\ref{fig:b2vsasq}. The
uncertainties also include the systematic uncertainty associated with the
particular choice for $R_{\rm min}$, estimated from the fit results obtained with
$R_{\rm min}\to R_{\rm min}\pm1a$, and from the unknown higher order terms,
estimated by the spread of the results from fits {\bf B} to {\bf D}. For the
continuum extrapolation we employ a function linear in $a^2$. To estimate
the systematics associated with this choice,
we compare the result to the ones obtained including only
the data with $(a/r_0)>0.2$. For error propagation of the other
systematic uncertainties, we perform all possible combinations of continuum
(and later also large-$N$) extrapolations and compute the individual systematic
uncertainties as described above. The continuum results for \btt{} are plotted in
Fig.~\ref{fig:b2vsNsq}.

We extrapolate \btt{} to the large-$N$ limit using a linear function in $1/N^2$
including data with $N>2$. As for the continuum extrapolation, we estimate the
systematic uncertainty by comparison to an extrapolation including data with
$N>3$. We obtain
\be
\label{eq:b2-largeN}
\bar{b}^{N\to\infty}_2=-0.0141(3)(15)(13)(9)(17) \,.
\ee
The first uncertainty is purely statistical, the second is the systematic one
associated with the higher order corrections, the third is the one for the choice
of $R_{\rm min}$, the fourth is the one of the continuum extrapolation and the fifth
the one of the large-$N$ extrapolation. The final result is also shown as the
black point in Fig.~\ref{fig:b2vsNsq}.

\begin{figure}[t]
 \centering
 \includegraphics[width=11cm]{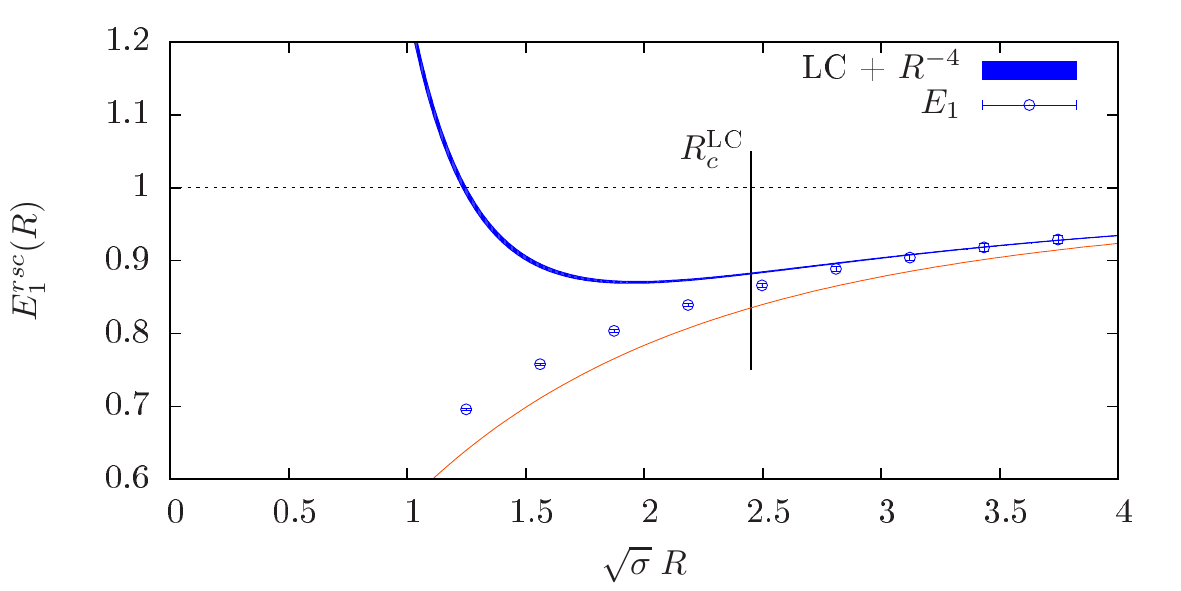}
 \caption{\label{fig:exc-state}
 Comparison between the EST prediction and the data for the first excited
 state $E_{1}$ for $\SU(2)$ at
 $\beta=5.0$. The red curve is the difference from the LC spectrum and
 the one labeled with `LC$+R^{-4}$' includes the BC term. The  vertical line
 with label `$R_c^{\rm LC}$' indicates the radius of convergence for the series
 expansion in $1/R$ of the first excited state of the LC spectrum.
 }
\end{figure}

Finally, we test the consistency with the excited states. In the EST
the energies are fully determined by
$\sigma$, $V_0$ and \btt{} up to higher order terms. We compare the EST
prediction for the first excited state with the data
for $\SU(2)$ at $\beta=5.0$ from Ref.~\cite{Brandt:2010bw} in Fig.~\ref{fig:exc-state}.
The data show good agreement with the curve up to $\sqrt{\sigma}R\approx3$, where
deviations due to higher order terms become visible. In fact, a naive fit including
a term of $\Ord(1/R^6)$ agrees with the data down to $\sqrt{\sigma}R\approx1.8$.
One can also extract \btt{} from the excited states~\cite{Brandt:2010bw}. The result,
which is in excellent agreement with the analysis of the potential,
is shown as the red band in Fig.~\ref{fig:exc-state}. We thus conclude that the results for
\btt{} are fully consistent with the excited state data. A more detailed comparison
is planned for the near future.

\section{Testing the presence of massive modes}

So far we have neglected possible massive-mode contributions. To test
how they would affect the results for \btt{} we repeat the analysis
from the previous section employing a fit function of the form
\be
\label{eq:mass-fit}
V(R) = E_{0}^{\rm EST}(R) - \frac{m}{2\pi} \sum_{k=1}^{\infty} \frac{K_1(2kmR)}{k}
- \frac{(d-2)(d-10)\pi^2}{3840 m \sigma R^4} +
\frac{\gamma^{(1)}_{0}}{\sqrt{\sigma^5} R^6}
+ \frac{\gamma^{(2)}_{0}}{\sigma^3 R^7} + V_0 \,.
\ee
In practice, the infinite sum is completely dominated by the first few terms,
so that it is sufficient to use the first 100 terms to reach machine precision. We perform
four different fits:
\begin{enumerate}
 \vspace*{-2mm}
 \item[{\bf F}] use $\sigma$, $V_0$, \btt{} and $m$ as free parameters, set
 $\gamma^{(1)}_{0}=\gamma^{(2)}_{0}=0$;
 \vspace*{-2mm}
 \item[{\bf G}] use $\sigma$, $V_0$, \btt{}, $m$ and $\gamma^{(1)}_{0}$ as free
 parameters, set $\gamma^{(2)}_{0}=0$;
 \vspace*{-2mm}
 \item[{\bf H}] use $\sigma$, $V_0$, \btt{}, $m$ and $\gamma^{(2)}_{0}$ as free
 parameters, set $\gamma^{(1)}_{0}=0$;
 \vspace*{-2mm}
 \item[{\bf J}] use $\sigma$, $V_0$ and $m$ as free parameters, set
 $\gamma^{(1)}_{0}=\gamma^{(2)}_{0}=\bt=0$.
 \vspace*{-2mm}
\end{enumerate}
The last fit is similar to fit {\bf E} above and checks whether the $R^{-4}$
term from Eq.~\refc{eq:pot-rigid} is already sufficient to describe the data. As before,
however, we find that fit {\bf J} needs much larger values of $R_{\rm min}$ and,
consequently, does not compare equally well to the data as the other fits. Unfortunately,
the results from fits {\bf G} and {\bf H} are not sufficiently precise, since the fits
contain two higher order terms. In the following we will thus only use fit {\bf F}.
Note, that this impedes the estimation of the systematic uncertainty due to
possible higher order terms.

\begin{figure}[t]
 \centering
 \includegraphics[width=12cm]{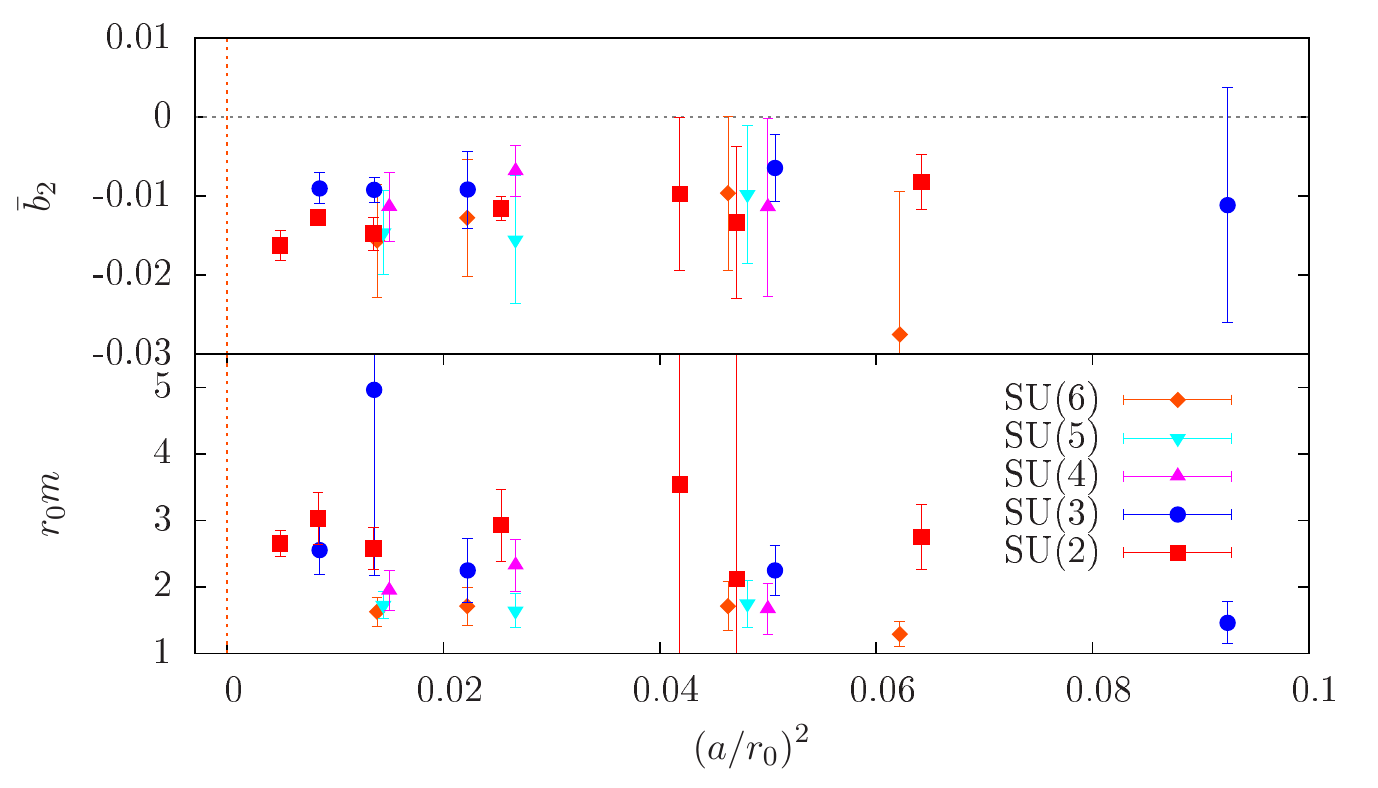}
 \caption{\label{fig:b2mvsasq}
 Results for the boundary coefficient \btt{} and the mass $m$ in units if $r_0$
 for the individual lattices versus the squared lattice spacing.
 }
\end{figure}

\begin{figure}[t]
 \centering
 \includegraphics[width=12cm]{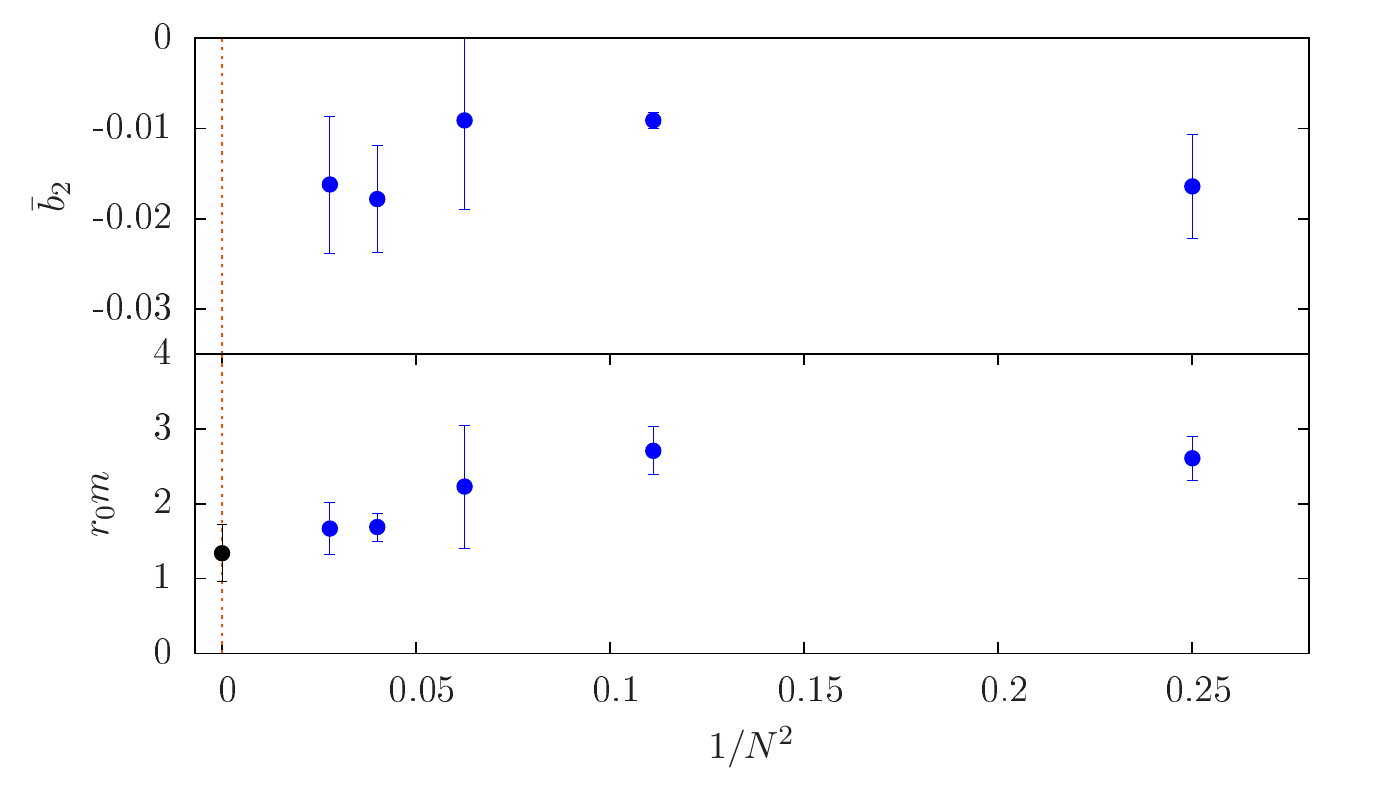}
 \caption{\label{fig:b2mvsNsq}
 Continuum results for the boundary coefficient \btt{} and the mass $m$
 versus $1/N^2$. The black point for $m$ is the result in the large-$N$ limit.
 }
\end{figure}

We show the results for \btt{} and $m$ in figure~\ref{fig:b2mvsasq}. Due to the 
additional $R^{-4}$ term, the results for \btt{} move closer to zero and,
in general, the uncertainties for \btt{} increase. The results for $m$ 
approach the continuum smoothly, enabling a linear continuum extrapolation.
The continuum results are shown in Fig.~\ref{fig:b2mvsNsq}. While the uncertainties
for \btt{} are too large to perform a reliable extrapolation to $N\to\infty$, the
results for $m$ allow for a reliable large-$N$ extrapolation and the final result is
\be
\label{eq:m-largeN}
r_0m^{N\to\infty}=1.34 (4)(8)(25)(27) \,, \quad \textnormal{or} \quad
\frac{m^{N\to\infty}}{\sqrt{\sigma}^{N\to\infty}}=1.1(4) \,.
\ee
Here the first uncertainty is purely statistical, the second is the systematic
uncertainty associated with $R_{\rm min}$, the third is the one of the continuum
extrapolation and the fourth is the one of the large-$N$ extrapolation. This result
can be compared to the one for the massive mode in 4d,
$m^{N\to\infty}/\sqrt{\sigma}^{N\to\infty}=1.713(4)$~\cite{Athenodorou:2017cmw}.
While $m$ appears to be somewhat smaller in 3d, one has to keep in mind, that the
4d result is not continuum extrapolated.
It is thus possible, that we are actually seeing a similar massive mode in 3 and 4
dimensions.

\begin{figure}[t]
 \centering
 \includegraphics[width=11cm]{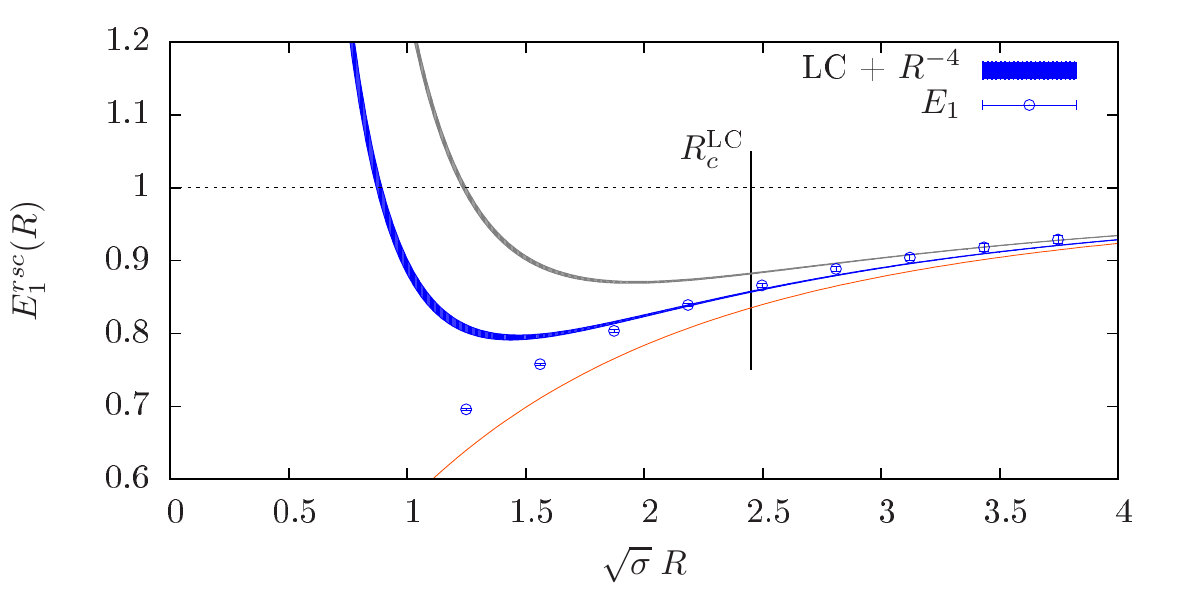}
 \caption{\label{fig:exc-state-m}
 Comparison between the EST prediction and the data for $E_{1}$
 for $\SU(2)$ and $\beta=5.0$ as in Fig.~8. The blue curve now includes \btt{} obtained
 from the analysis with massive modes, while the gray curve is the one from Fig.~8.
 }
\end{figure}

We can again compare the result for \btt{} with the $\SU(2)$, $\beta=5.0$
results for the excited states. The comparison is shown in Fig.~\ref{fig:exc-state-m}.
The prediction (blue curve) lies below the data even at large $R$.
Naively, this implies a contradiction with the excited state data. However, the
excited state contributions of the massive mode are unknown,
rendering the comparison incomplete.

\section{Conclusions}

In this proceedings article I provided an update on my studies of the energy levels
of the (open) flux tube and their comparison to the EST in 3d $\SU(N)$ gauge
theories. I obtain continuum and large-$N$ extrapolated results for the EST
parameters with full control over the relevant systematic effects. The
large-$N$ extrapolated results are given in
Eqs.~\refc{eq:sig-r-res},~\refc{eq:b2-largeN} and~\refc{eq:m-largeN}.
I have also shown that the leading order correction to the LC spectrum is indeed
of $\Ord(R^{-4})$. The main uncertainty for \btt{} concerns the presence or
absence of massive modes. In both cases, however, \btt{} remains
non-vanishing at finite $N$ and, likely, also in the large-$N$ limit (cf.
Fig.~\ref{fig:b2mvsNsq}). It would be interesting to obtain a prediction for
the contribution of massive modes to the excited states. This could potentially
help to either rule out or confirm the presence of massive modes and might even
enable the discrimination of massive mode and rigidity contributions. It is
intriguing to see that the result for $m$ is in good agreement with the masses
found in~\cite{Athenodorou:2017cmw}, providing a hint for a similar origin of
the massive modes in 3d and 4d. In the large-$N$ limit the KKN prediction
agrees with the lattice result up to 1.6\%, which is about a factor of two
further away than previous findings~\cite{Lucini:2002wg,Bringoltz:2006gp}.

\end{document}